\documentclass[letterpaper]{article}

\usepackage[T1]{fontenc}

\usepackage{geometry}
\usepackage{setspace}
\usepackage{hyperref}
\usepackage{pdfpages}

\usepackage[backend=biber,natbib=true,hyperref=true,style=chem-angew,articletitle=true,doi=true]{biblatex}
\usepackage{graphicx}
\usepackage{float}
\newfloat{scheme}{htbp}{los}
\floatname{scheme}{Scheme}
\floatname{chart}{Chart}
\newfloat{graph}{htbp}{loh}

\usepackage{chemformula} 
\usepackage[version = 4]{mhchem} 

\usepackage{authblk}
\author[1]{Alba de las Heras*}
\affil[1]{Max Planck Institute for the Structure and Dynamics of Matter \& Center for Free-Electron Laser Science, Hamburg 22761, Germany}
\author[2]{Ofer Neufeld*}
\affil[2]{Technion Israel Institute of Technology, Faculty of Chemistry, Haifa 3200003, Israel}
\author[1,3]{Angel Rubio}
\affil[3]{Center for Computational Quantum Physics, The Flatiron Institute, New York 10010, USA}

\title{Pulse-duration-sensitive high harmonics and attosecond locally-chiral light from a  chiral topological Weyl semimetal}
\date{*Emails: alba.de-las-heras@mpsd.mpg.de,  ofern@technion.ac.il, angel.rubio@mpsd.mpg.de}

\begin{document}

\maketitle

\begin{abstract}
  High harmonic generation (HHG) in solids results from an interplay between intraband acceleration and electron-hole recombination driven by a high-intensity laser pulse. Here, we theoretically reveal that the driving pulse duration can play a major role in extending HHG to higher photon energies by promoting higher conduction band excitations. The effect is present in a conventional semiconductor as Si, restricted in a large-gap insulator as MgO, and most prominent in RhSi, a prototypical chiral Weyl semimetal presenting numerous band crossings. Further, we elucidate the HHG selection rules in RhSi required for the synthesis of attosecond locally chiral light. The chiral crystal structure enables the generation of a local 3D electric field exhibiting an asymmetric instantaneous torsion on attosecond timescales. A pronounced circular dichroism emerges when the driving helicity is either aligned with or opposite to the crystal handedness. Our findings motivate future experiments in chiral Weyl semimetals to track high-energy band crossings and in-situ locally chiral light, paving the way for chiral compact light sources and light-wave driven topological electronics.
\end{abstract}

\section*{Keywords}
Attosecond science; Quantum topological materials; Chirality; High harmonic generation; Ultrafast electron dynamics



\section{Introduction}

The interaction of high-intensity laser pulses with matter provides valuable insight into electron dynamics and enables the development of high-frequency sources in the extreme ultraviolet and soft X-ray regions~\cite{Krausz2009,Popmintchev2010}. The coherent nature of high harmonic generation (HHG) is a crucial ingredient allowing the synthesis of attosecond pulses as well as the traceability of ultrafast dynamics~\cite{Paul2001,Hentschel2001,Kling2008}. Dilute gases composed of atomic or molecular species have long been the standard targets for HHG, until the first observation in a bulk crystal~\cite{Ghimire2011}. Nowadays, HHG from quantum topological materials is an extremely active research line, covering Dirac~\cite{Yoshikawa2017,Kovalev2020,Garcia-Cabrera2024} and Weyl semimetals~\cite{Lv2021,Avetissian2022,Bharti2023,Bharti2024,Ominato2025,Liu2025,Zhang2024,Yao2025}, as well as topological insulators~\cite{Schmid2021,Bai2021,Baykusheva2021}. Chiral Weyl semimetals are particularly attractive due to the combined handedness in the crystal lattice and in the electronic Weyl fermions~\cite{Chang2017,Hasan2021,Yan2024}. The potential of HHG from chiral topological materials presenting dual electronic and structural chirality has not yet been exploited. 

The high harmonic spectra from gases and solids are similarly characterized by the presence of nonperturbative high harmonic plateaus ending at a cutoff energy~\cite{Li1989,Ghimire2011}. However, the well-known picture of HHG in atoms and molecules, based on the three steps of i) tunnel ionization, ii) acceleration during the electron real-space excursion and iii) recombination~\cite{Schafer1993,Corkum1993}, is more complex in solids. HHG in solids results from intraband and interband contributions~\cite{Vampa2014}. Intraband HHG originates from the nonlinear electron motion within a band, whereas the interband pathway requires i) electron excitation into a conduction band, ii) displacement in momentum space and iii) electron-hole recombination~\cite{Ghimire2019}.
The understanding of interband and intraband contributions in solid-state HHG enables the retrieval of quantitative information of the band structure~\cite{Luu2015,Vampa2015,Ndabashimiye2016,Tancogne-Dejean2017bs,Langer2018,Uzan-Narovlansky2022,Tyulnev2024}, crystal symmetries~\cite{You2017,Uzan-Narovlansky2023}, and sub-cycle carrier dynamics~\cite{Schubert2014,Uzan-Narovlansky2024}. 
Overall, the energy cutoff in HHG is typically ruled by interband dynamics, so that the maximum photon energy is given by the maximum electron-hole energy difference upon recombination~\cite{Wu2015}. Multiple plateaus are known to emerge due to high-energy transitions from higher conduction bands~\cite{Wu2015,Hawkins2015,Wu2016,Ikemachi2017,Ndabashimiye2016,You2017MgO,Allegre2025}. A subtle parameter in HHG from solids is the driving pulse duration, which has been predicted in 1D models to be crucial in the emergence of secondary plateaus~\cite{Ikemachi2017}. Nevertheless, to the best of our knowledge, such pulse duration dependence on solid-state HHG has not yet been reported from experiments or simulations.

Crystals offer an additional opportunity to imprint their symmetries on the high-frequency coherent emission, which departs from the conventional approach of tailoring the driving field to tune and control the properties of HHG~\cite{delasHeras2024, DelasHeras2022, Neufeld2019}. For instance, high harmonic emission can be naturally produced with a high ellipticity in solids~\cite{Saito2017,Tancogne-Dejean2017,Klemke2019}, as an alternative to the schemes proposed in gas targets~\cite{Hickstein2015,Fan2015,Huang2018,Brooks2025}. Equally important, the absence of certain symmetries enables richer phenomena, such as the circular photogalvanic effect in materials lacking inversion symmetry~\cite{Rees2020,Ni2021,Ma2022,Neufeld2021,Lesko2025}. The absence of mirror symmetry, known as chirality, is vividly studied in molecular systems ~\cite{Neufeld2019chiral,Wanie2024, Habibovi2024,Han2025,Chen2024,Haase2026} because enantiomer sensitivity is paramount in biomedical applications and in catalytic processes for chemical design. Similarly, HHG can probe the structural chirality of crystals\cite{Chen2020,Heinrich2021}. However, chiral crystals have not been investigated as a way to imprint local chirality in HHG. The key concept of locally chiral light is to engineer a non-superimposable mirror-image electric field microscopically for a more efficient chiral interaction with matter\cite{Ayuso2019,Neufeld2020,Ayuso2022,Rego2023,Mayer2024}. Remarkably, matter systems are sensitive to the handedness of locally chiral light irrespective of their spatial scale\cite{Ayuso2019}. Still, the generation and characterization of locally chiral light remains experimentally challenging, due to the requirements of high spectral control and near-field detection, and its interaction and emergence in chiral solids has not been explored thus far.     

Here, we investigate HHG in RhSi ---a well-established chiral Weyl semimetal\cite{Chang2017,Hasan2021,Yan2024}--- using time-dependent density-functional theory (TDDFT). The bifold aim is (i) to unveil the role of pulse duration in solid-state HHG, and (ii) to demonstrate that the structural chirality maps into locally chiral HHG. Our \textit{ab initio} calculations show that the duration of the driving laser pulse substantially influences the quantum pathways of electron excitation to higher conduction bands and, in consequence, the cutoff energy in HHG. The effect is paramount in RhSi due to its high multiband coupling, moderate in Si, and severely suppressed in the large-gap insulator MgO. Additionally, in RhSi, we identify HHG selection rules to synthesize locally chiral light. By exploiting the strong-field interaction of a circularly polarized driver with the chiral crystal structure, it is possible to engineer a 3D local electric field presenting an asymmetric instantaneous torsion on attosecond timescales.  Such locally chiral polarization varying on hundreds of attoseconds sets a new route to induce and probe attosecond chiral excitations within dipolar interactions. Overall, our work establishes chiral Weyl semimetals as a promising material family for observing and manipulating out-of-equilibrium coherent phenomena, eventually advancing compact extreme-ultraviolet sources and ultrafast light-wave driven electronics. 

\section{Results}
\subsection{Pulse duration dependence of high harmonic generation from bulk crystals}
\begin{figure}[H]
	\centering
    \includegraphics[width=0.99\textwidth]{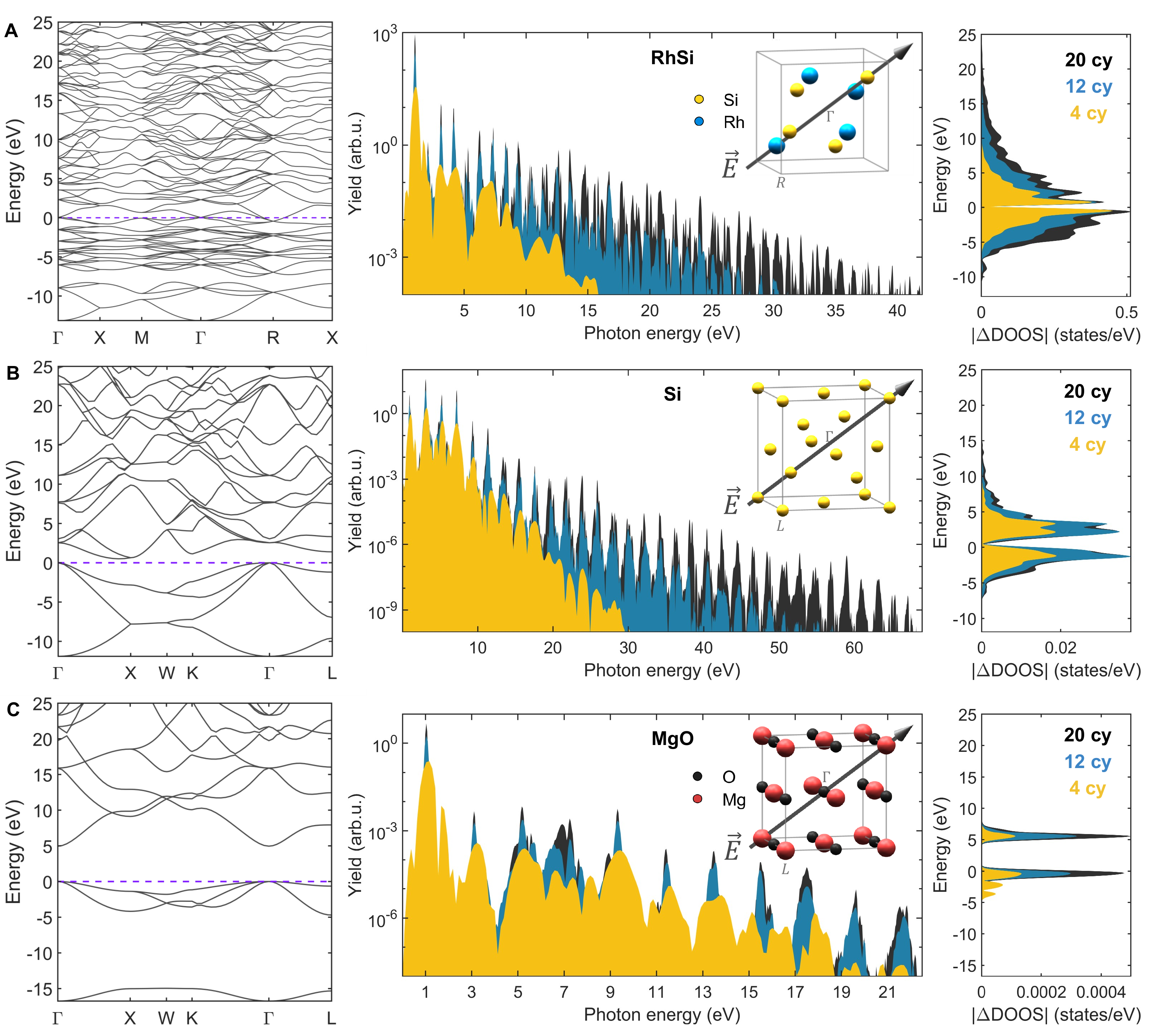} 
	\caption{\textbf{Pulse duration dependence of high harmonic generation in bulk crystals.}
		Computed band structure (left), high harmonic spectrum (center) and change of the density of occupied states (right) for three different materials: (\textbf{A}) RhSi, (\textbf{B}) Si and (\textbf{C}) MgO. The Fermi energies are indicated at 0 eV (dashed violet line). The full duration of the driving laser pulse is set to 4 (yellow), 12 (blue) and 20 (black) optical cycles. In RhSi and Si, longer durations of the driving pulse are associated with the occupation of higher-energy states and a substantial extension of the emission towards higher photon energies. The insets show the unit cell and the direction of the linearly polarized electric field.}
	\label{fig:hhspectrapulseduration} 
\end{figure}

We first consider a laser pulse with a central photon energy of ${\sim}1$~eV and linear polarization along the [111] crystal direction. In Fig.~\ref{fig:hhspectrapulseduration}, we compare driving full pulse durations of 4, 12, and 20 optical cycles, respectively shown in yellow, blue and black. Each panel presents the
band structure, the high harmonic spectrum, and the change of the electronic density of occupied states, $|\Delta \text{DOOS}|$ in Eq.~\ref{eq:doos}, for (A) RhSi, (B) Si and (C) MgO. The inset shows the crystal unit cell and the geometry of the incident electric field. Following the usual practice, the origin of the energy axes is shifted to the Fermi energies. To produce comparable HHG responses, the driving peak intensity is set to 1 ~TW/cm$^{2}$ in RhSi and Si, whereas 3~TW/cm$^{2}$ is chosen for MgO due to its larger band gap.

Figure~\ref{fig:hhspectrapulseduration} shows an extreme sensitivity of the energy cutoff of HHG in RhSi and Si to the pulse duration. In Si, the cutoff is extended from ${\sim}25$ to ${\sim}60$~eV when increasing the pulse duration from 4 to 20 cycles, while keeping constant the peak intensity and other driving parameters. The harmonic yield follows a characteristic multiplateau behaviour, presenting abrupt stepped decays at ${\sim}7$, 26, 43, and 59~eV. Consequently, the highest plateau (in the energy range from 43-59 eV) is emitted with a reduction in the harmonic yield of 8 orders of magnitude, compared to the standard plateau (at $<7$~eV). In RhSi, the cutoff extension expands from ${\sim}7$ to ${\sim}35$~eV. Noticeably, the spectra exhibit a smoother decay of the harmonic yield, covering a harmonic yield reduction of 3 orders of magnitude. This establishes RhSi as a promising material for observing the energy cutoff extension in experiments. While the energy cutoff is expected to depend on the driving photon energy and peak field amplitude\cite{Liu2017,Ghimire2011}, the influence of the pulse duration implies a progressive electron ascension through the conduction bands during the interaction with the laser pulse~\cite{Ikemachi2017}.  

As an opposite case, the high harmonic spectrum in MgO (Fig.~\ref{fig:hhspectrapulseduration}C) does not present a notorious dependence of the energy cutoff on the pulse duration. We recognize the standard effects of longer pulses, such as the harmonic peak narrowing and the yield enhancement coming from the periodicity of more recombination events. In the following, we aim for more insights into the origin of the cutoff extension and the material characteristics that favour or hinder this mechanism. To evaluate the electron rearrangement induced by the laser pulse, we define the change in the density of occupied states by  
\begin{equation}
\label{eq:doos}
    |\Delta \text{DOOS}(E)|=\left|\sum_{k,n}  \left[f_{k,n}(t=\tau_\text{pulse})-f_{k,n}(t=0)\right] \, w_k \, \delta(E - E_{k,n})\right|,
\end{equation}
where $f_{k,n}(t)$ are the occupation coefficients of each eigenstate at time $t$, $w_k$ the k-point weights, $E$ the energy variable, $E_{n,k}$ the eigenvalues, $\tau_\text{pulse}$ the full duration of the driving laser pulse and $t=0$ the initial time before the laser pulse. The function $|\Delta \text{DOOS}(E)|$ is plotted in the right column of Fig.~\ref{fig:hhspectrapulseduration} for different driving pulse durations in RhSi, Si and MgO. Longer pulses promote electron population to high-energy states in RhSi and Si, but not in MgO. This suggests that \textit{``excited ladder electrons"} (i.e. electrons occupying high-energy conduction bands after traveling through band crossings or narrow band gaps) are responsible for the cutoff extension that we observe in the high harmonic spectra of RhSi and Si. 

The occupations in MgO show that the valence bands are only coupled to the  first conduction band. This dismantles the possibility of increasing the energy difference between recombining electrons and holes that is required for higher-order harmonic plateaus. The large band gap of MgO prevents a strong coupling to higher energy states, since the dipole matrix elements are inversely proportional to the energy difference of the transition. It requires substantially higher intensities to observe a secondary plateau\cite{Allegre2025}. Conversely, the intricate network of band crossings observable in the band structure of RhSi favours the multiband coupling and the ascent of electrons to high-energy states even for lower driving intensities. This is in consonance with the higher harmonic yield in RhSi at high photon energies. Silicon can be interpreted as an intermediate situation, where the smaller band gap allows the multiband coupling but it is not as favoured as in RhSi. This is also compatible with the reduced yield obtained in the HHG spectra in Si.  

\begin{figure}[H]
	\centering
    \includegraphics[width=0.85\textwidth]{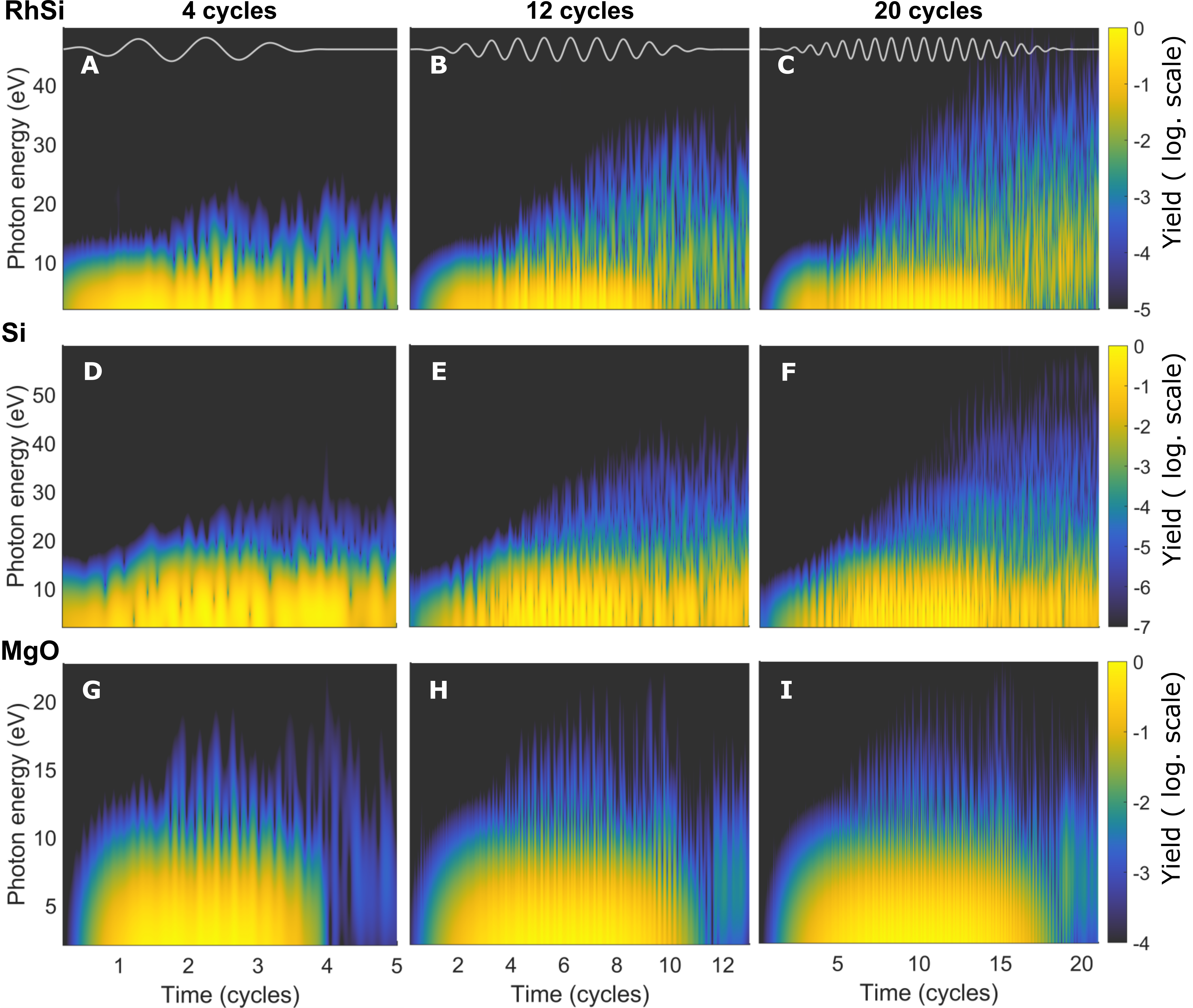} 
	\caption{\textbf{Time-resolved spectrogram of the high harmonic emission in RhSi, Si and MgO.} The emission of high photon energies requires long interaction times in RhSi and Si. In both crystals, the energy cutoff increases with the interaction time. In contrast, the emission in MgO reaches similar photon energies independently of the pulse duration. The driving pulses of 4 (left), 12 (center) and 20 optical cycles (right) are shown in grey lines at the top.}
	\label{fig:spectrogram} 
\end{figure}

Even if high-energy excitation pathways are allowed by the material characteristics and the laser parameters, the occupation of high-energy states requires several interaction cycles\cite{Ikemachi2017}. The mechanism of excited ladder electrons is cumulative, so the population of higher-energy states increases for longer driving pulses. For a better insight into the progressive build-up of HHG, we continue the analysis of the emitted radiation. In Fig.~\ref{fig:spectrogram}, the time-resolved spectrogram of HHG in RhSi, Si and MgO provides information about the photon energy emission as a function of time. Again, we compare driving pulse durations of 4 (left column), 12 (center) and 20 (right column) optical cycles. On the one hand, in RhSi and Si, higher photon energies are emitted as the number of optical cycles increases. After the initial cycles, the energy cutoff scales linearly with the interaction time (see Figs.~\ref{fig:spectrogram}B,C,E,F). The emission of high photon energies persists even at the end of the laser pulse, until the eventual vanishing due to subsequent decoherent effects not included in the simulations.

On the other hand, the emission in MgO shows a similar behavior for all pulse durations (Figs.~\ref{fig:spectrogram}G-I), which resembles the usual HHG in atoms and molecules. In MgO, the photon energies around the energy cutoff are emitted around the central times of the pulse when the peak intensity is higher. This evidences the different HHG emission from a few-band-coupling in MgO or a high-multiband-coupling in Si and RhSi. The sensitivity of HHG to interband coupling enables tracking information about the excited electron dynamics. Concretely, the energy cutoff provides access to the maximum electron-hole energy difference upon recombination. Thus, with the time-resolved spectrogram analysis, we retrieve the time required for high-energy electron-hole recombination. The duration of the laser-matter interaction determines the excited states that electron pathways can reach, and typically, the excited electron population increases during every laser cycle. The optimal duration for extending the cutoff needs to be experimentally investigated, looking for a trade-off between electron excitation and decoherence while satisfying macroscopic phase-matching conditions. Theoretically, even higher cutoff energies can be obtained by increasing the driving intensity (see Supplemental Material\cite{methods}).    

\subsection{Symmetry selection rules and attosecond locally chiral light in RhSi}

\begin{figure}[H]
	\centering
\includegraphics[width=0.99\textwidth]{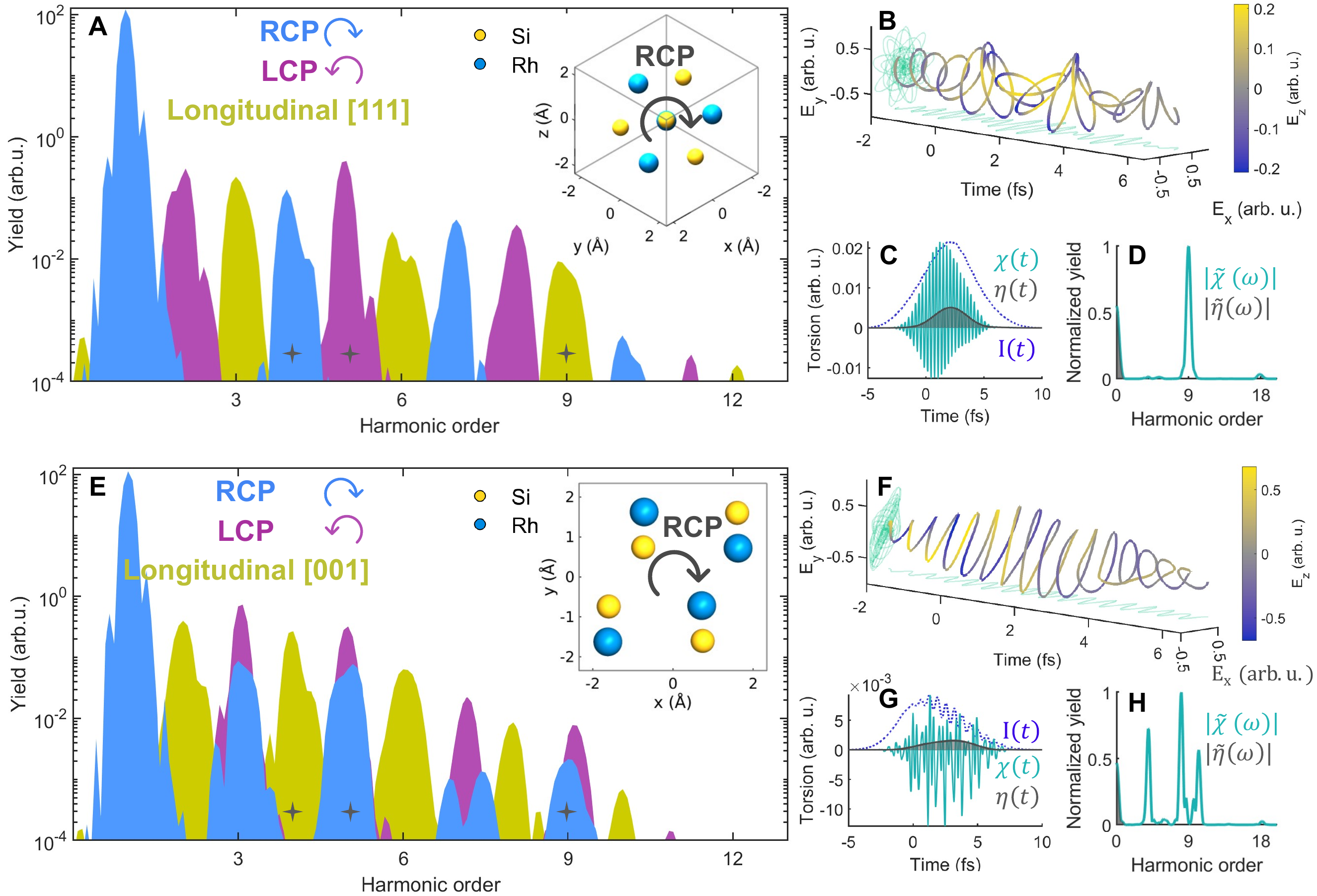} 
	\caption{\textbf{Chiral high harmonic generation in RhSi driven by a RCP pulse.} The interaction of RhSi with a right circularly polarized (RCP) driver results in a complex emission with in-plane and out-of-plane polarization components. (\textbf{A}) The high harmonic spectrum driven by a RCP pulse in the (111) crystal plane contains alternate harmonic orders of counterrotating circular polarization (RCP in blue and LCP in violet) followed by a longitudinal component (yellow). The interaction geometry is represented in the inset. The harmonic orders 3, 4, and 9 selected for the spectral synthesis are indicated with grey stars. (\textbf{B}) Time evolution of the 3D synthesized electric field. Its local chirality is quantified via the instantaneous torsion (light blue) and the torsion envelope (shaded area) plotted in (\textbf{C}) time and (\textbf{D}) frequency domains. The dotted line represents the electric field intensity, $I(t)$, in arbitrary units as a reference. (\textbf{E-H}) Equivalent plots are shown for a RCP driving laser pulse in the (001) crystal plane.}
	\label{fig:attochiral} 
\end{figure}

In this section, we examine the dynamical symmetries of HHG~\cite{Alon1998,Neufeld2019} in the chiral Weyl semimetal RhSi. For a linearly polarized driver along the chiral crystal axis [111], the emission is linearly polarized along the same direction. To exploit the chirality of the crystalline structure in HHG, we change the driving polarization to circular. Figure~\ref{fig:attochiral} shows the HHG and the synthesized electric field for two different crystal orientations. We set a driving photon energy of 1.55~eV,  a pulse duration of 8 cycles, and a peak intensity of 1~TW$/$cm$^2$.

If a right circularly polarized (RCP) driving laser pulse propagates along the chiral axis [111], the high harmonic spectrum (Fig.~\ref{fig:attochiral}A) shows triplets of harmonic orders with the same helicity (RCP in blue), opposite helicity ---i.e. left circular polarization (LCP) in violet--- and longitudinal polarization (yellow). This selection rule is rarely found in HHG, since it requires a three-fold rotational symmetry in tandem with a lack of mirror symmetry transverse to the rotation axis\cite{Neufeld2019} (whereas gas-phase systems are typically isotropic and therefore exhibit the mirror plane). We denote it by $C_3^{(\pm)}$, where $(\pm)$ labels the right- and left-handed enantiomorphs associated with the crystal lattice.  
The absence of inversion symmetry and mirror planes is a direct feature of the B20 cubic chiral crystal structure (space group 198), enabling the emergence of longitudinally polarized harmonics (the $3n$ components in the triplets of integer $n$). These longitudinally polarized harmonics do not propagate in free space according to Maxwell's equations due to phase-matching constraints. They constitute evanescent fields that can be detected in the near field~\cite{Zhang2013,Poulikakos2018}, and be used for imaging and spectroscopic applications\cite{Fish2022,Li2017}. Future efforts can be directed to the design of noncollinear configurations satisfying the $C_3^{(\pm)}$ dynamical class symmetry while enabling the propagation of longitudinal components (e.g. as proposed in refs.~\cite{Neufeld2019chiral,Ayuso2019}). Nonetheless, in the following, we focus on analyzing the microscopic 3D electric field, including both in-plane and out-of-plane components, which can be used in-situ for spectroscopies in the chiral crystal and for the synthesis of locally-chiral light.  

The harmonic triplets resulting from the dynamical symmetry class $C_3^{(+)}$ in Fig.~\ref{fig:attochiral}A provide a main element for the local chiral light synthesis: harmonic orders with three different orthogonal polarizations. The frequency of the longitudinally polarized harmonics can be matched with the frequency sum of the coplanar LCP and RCP components to yield a nonvanishing average chirality in the local electric field (i.e., locally chiral light)\cite{Ayuso2019}. 
The electric field resulting from the coherent superposition of harmonic orders 3, 4 and 9 is displayed in Fig.~\ref{fig:attochiral}B in the temporal domain. The electric field is normalized, and the longitudinal component is represented with the colormap. The projected views are also plotted (in light blue) as a reference. 
 
To quantify the light's local chirality, we define the instantaneous time-dependent torsion
\begin{equation}
    \label{eq:torsion}
    \chi(t)=\mathbf{E}(t)\cdot \left(\dfrac{d\mathbf{E}(t)}{dt}\times\dfrac{d^2\mathbf{E}(t)}{dt}\right).
\end{equation}
The pseudoscalar $\chi(t)$ describes the noncoplanar twist of the trajectory traced by the local electric field vector in time and electric-field space. 
$\chi(t)$ evaluates the instantaneous chirality of the local electric field vector $\mathbf{E}(t)$. We consider a normalized electric field, so that $\chi(t)$ is in normalized units (n.u.). Figure~\ref{fig:attochiral}C shows the ultrafast evolution of the instantaneous torsion (light blue line), the torsion envelope, $\eta(t)$ (shaded area), and the intensity of the local field, $I(t)$ (dotted line, in arbitrary units). Notably, the torsion envelope, $\eta(t)$, evidences the asymmetry in the instantaneous torsion $\chi(t)$. 
The time integral $\zeta=\int \chi (t)dt$ indicates the net torsion carried by the electric field (in our case, $\zeta=0.3679$~n.u.). While an electric field can be instantaneously locally chiral if $\chi(t)\neq0$, a net local chirality requires $\zeta\neq0$. The definitions of $\chi(t)$, $\eta(t)$ and $\zeta$ can be applied to quantify the instantaneous, period-averaged and net chirality of any local electric field, independently of the spectral components or the complexity of its temporal waveform.     

Further information can be gained with the spectral decomposition of the instantaneous torsion (see Fig. ~\ref{fig:attochiral}D). Denoting by $\omega_0$ the central driving frequency, the main frequency of the local electric field torsion is $9\omega_0$, which corresponds to a period of $296$ as. Smaller peaks are observed at $4\omega_0$ and $5\omega_0$ (with periods of 666 and 533 as), as expected from the synthesized harmonic orders. Besides, the torsion envelope is associated with the low-frequency peak (shaded area in Fig. ~\ref{fig:attochiral}D). These results indicate that HHG from chiral Weyl semimetals can be a useful tool for attosecond chiral spectroscopy and for triggering chiral dynamics by exploiting both high and low-frequency torsional components.

Panels~\ref{fig:attochiral}E-H show HHG and the electric field analysis for a different crystal orientation. When the RCP laser pulse is contained in the (001) crystal plane, the HHG spectrum is governed by a in-plane two-fold rotational symmetry while still breaking transverse mirror symmetry, $C_2^{(\pm)}$. This restricts the in-plane components to odd harmonic orders, and longitudinal polarization to even harmonics (Fig.~\ref{fig:attochiral}E). The even harmonics present a high ellipticity because the RCP and LCP components do not exhibit the same amplitude, which is caused by the screw crystal axis [001]. A similar effect can arise in specific harmonic orders in other materials due to anomalous ellipticity behaviors originating from the bandstructure\cite{Tancogne-Dejean2017}. As in the previous configuration, we synthesize the electric field by filtering harmonics 4, 5 and 9. The resulting electric field is shown in Fig.~\ref{fig:attochiral}F. The selection rule $C_2^{(+)}$ leads as well to a local 3D electric field exhibiting instantaneous torsion (light blue line in Fig.~\ref{fig:attochiral}G) varying on attosecond scales. However, the instantaneous torsion in Fig.~\ref{fig:attochiral}G reaches lower peak values and is more symmetric than in the $C_3^{(+)}$ selection rule (Fig.~\ref{fig:attochiral}C). The increase in the symmetry translates into a reduction of the torsion envelope and the net torsion, which in this case is $\zeta=-0.1402$~n.u. The negative sign indicates an opposite handedness in the torsion. The spectral components of the torsion (Fig.~\ref{fig:attochiral}H) show a more complex behavior, but still we observe both low and high frequency components. 

In summary, the $C_2^{(+)}$ selection rule also allows for the synthesis of locally chiral light carrying an asymmetric instantaneous torsion varying on attosecond scales, a non-vanishing torsion envelope, and net torsion. Still, the $C_3^{(+)}$ selection rule is beneficial for higher net local chirality. In the macroscopic picture, the total synthesized electric field using $C_3^{(\pm)}$ or $C_2^{(\pm)}$ selection rules is also expected to be globally chiral\cite{Ayuso2019}, since it is constituted by locally chiral electric fields with uniform handedness. More results considering high or low control in the spectral filtering, as well as the circular dichroism manifested when using the opposite driving helicity, are included in the Supplemental Material\cite{methods}. 
The emergence of local electric fields with instantaneous torsion occurs robustly in the light-driven chiral Weyl semimetal, which should be applicable for emerging attosecond locally-chiral sources with tunable temporal helicity.

\section{Conclusions}
In conclusion, our TDDFT calculations of HHG in RhSi and Si show a significant cutoff extension by increasing the driving pulse duration. We attribute this observation to the population of high conduction bands, which leads to higher electron-hole energy differences and thus, higher cutoff energies. This pathway of ``excited ladder electrons" is particularly favoured on the semimetal RhSi due to the large number of band crossings and the strong multiband coupling. In this material, the multiplateau extension merges into the main high harmonic plateau reaching higher photon energies, whereas the yield of the higher-order plateaus is weaker in Si and severely suppressed in MgO. The large band gap in MgO requires a substantially higher intensity in order to observe a secondary plateau~\cite{Allegre2025}, and whether electrons can reach higher conduction bands remains an incognita in this material. The high cutoff energies obtained in RhSi and Si surpass the current limit of 50~eV measured in MgO~\cite{Allegre2025} (see also Supplemental Material~\cite{methods}). The optimal parameters for extending the cutoff require further experimental investigation, looking for a trade-off between electron excitation and decoherence while satisfying macroscopic phase-matching conditions. Our results definitely broaden the search for optimal materials for HHG, which is a crucial issue for compact and coherent extreme-ultraviolet solid-state sources. Moreover, the mechanism of excited ladder electrons constrains the validity of few-band theoretical models, since they would fail to capture such high-excitation pathways in materials exhibiting strong coupling among multiple bands.

We also demonstrate that the chiral crystal structure of RhSi maps into locally chiral HHG\cite{Ayuso2019}. The longitudinally-polarized harmonics are often ignored since they only arise in the absence of transverse mirror symmetry, and they do not propagate to the far field. However, here, longitudinally polarized harmonics are a crucial component for the synthesis of local electric fields exhibiting a 3D chiral polarization varying on attosecond scales. The local electric fields retrieved from our TDDFT calculations present both low- and high-frequency torsional components. The instantanous torsion and the nonvanishing ultrashort torsion envelope can both be exploited for a more efficient dipolar chiral probe. A strong circular dichroism is observed for a driving propagation direction along the chiral crystal axis, depending on whether the driving-field helicity is opposite to or aligned with the handedness of the crystal lattice (see Supplemental Material~\cite{methods}). We find that the $C_3^{(+)}$ symmetry class, associated with a three-fold rotational symmetry plus lack of transverse mirror symmetry, is advantageous for the locally chiral light synthesis over the $C_2^{(+)}$. The reason is that the $C_3^{(+)}$ enables a spectral separation into triplets of harmonics with orthogonal polarizations, whereas the $C_2^{(+)}$ mixes counterrotating circular polarization in the even harmonic orders. Among the chiral crystals compatible with the $C_3^{(+)}$, the B20 cubic crystal structure (space group 198) is a special choice because it allows the emergence of multifold fermions\cite{Hasan2021} or magnetic skyrmions\cite{Muhlbauer2009}. 
Overall, our findings pave the way for future experiments measuring exotic attosecond light pulses, chirality, and high-energy excitation pathways through solid-state high harmonic generation. These are appealing prospects for compact EUV sources, sensitive enantiomer detection, and ultrafast light-wave-driven electronics.

\section*{Acknowledgements}

This work was supported by the European Research Council (ERC-2024-SyG-101167294; UnMySt), the Cluster of Excellence Advanced Imaging of Matter (AIM), Grupos Consolidados y Alto Rendimiento UPV/EHU, Gobierno Vasco (IT1453-22) and SFB925.  
We acknowledge support from the Max Planck-New York City Center for
Non-Equilibrium Quantum Phenomena. The Flatiron Institute is a division of the Simons Foundation. A.H. acknowledges funding from the Alexander von Humboldt Foundation. O.N. gratefully acknowledges support from the Technion Helen Diller Quantum Center and RBNI Nevet programs, as well as the Vatat Quantum Science and Technology Fellowship.

\section{Author contributions}
A.H. performed the numerical calculations and drafted the manuscript. All authors contributed to the study design, scientific discussions, and refinement of the manuscript.

\section*{Supporting information}
\begin{itemize}
    \item Methods
    \item Complementary results about:\\
(i) The HHG extension with the pulse duration for a higher driving intensity \\
(ii) The role of spectral filtering in locally chiral HHG\\
(iii) Locally chiral HHG driven by a left-circularly polarized driver\\
\end{itemize}

\printbibliography

@Misc{methods,
  note = {Methods and complementary results are available as supplementary material},
}

@article{Krausz2009,
author = {Krausz, Ferenc and Ivanov, Misha},
doi = {10.1103/RevModPhys.81.163},
issn = {0034-6861},
journal = {Reviews of Modern Physics},
month = {feb},
number = {1},
pages = {163--234},
publisher = {American Physical Society},
title = {{Attosecond physics}},
volume = {81},
year = {2009}
}

@article{Popmintchev2010,
author = {Popmintchev, Tenio and Chen, Ming-Chang and Arpin, Paul and Murnane, Margaret M. and Kapteyn, Henry C.},
doi = {10.1038/nphoton.2010.256},
issn = {1749-4885},
journal = {Nature Photonics},
month = {dec},
number = {12},
pages = {822--832},
publisher = {Nature Publishing Group},
title = {{The attosecond nonlinear optics of bright coherent X-ray generation}},
volume = {4},
year = {2010}
}

@article{Kling2008,
author = {Kling, Matthias F. and Vrakking, Marc J.J.},
doi = {10.1146/annurev.physchem.59.032607.093532},
issn = {0066-426X},
journal = {Annual Review of Physical Chemistry},
month = {may},
number = {1},
pages = {463--492},
publisher = {Annual Reviews},
title = {{Attosecond Electron Dynamics}},
volume = {59},
year = {2008}
}

@article{Schafer1993,
author = {Schafer, K. J. and Yang, Baorui and DiMauro, L. F. and Kulander, K. C.},
doi = {10.1103/PhysRevLett.70.1599},
issn = {0031-9007},
journal = {Physical Review Letters},
month = {mar},
number = {11},
pages = {1599--1602},
pmid = {10053336},
publisher = {American Physical Society},
title = {{Above threshold ionization beyond the high harmonic cutoff}},
volume = {70},
year = {1993}
}

@article{Corkum1993,
author = {Corkum, P. B.},
doi = {10.1103/PhysRevLett.71.1994},
issn = {0031-9007},
journal = {Physical Review Letters},
month = {sep},
number = {13},
pages = {1994--1997},
pmid = {10054556},
publisher = {American Physical Society},
title = {{Plasma perspective on strong field multiphoton ionization}},
volume = {71},
year = {1993}
}

@article{Ghimire2011,
   author = {Shambhu Ghimire and Anthony D. Dichiara and Emily Sistrunk and Pierre Agostini and Louis F. Dimauro and David A. Reis},
   doi = {10.1038/nphys1847},
   issn = {17452481},
   issue = {2},
   journal = {Nat. Phys.},
   pages = {138-141},
   publisher = {Nature Publishing Group},
   title = {Observation of high-order harmonic generation in a bulk crystal},
   volume = {7},
   year = {2011}
}

@article{Ghimire2019,
   author = {Shambhu Ghimire and David A. Reis},
   doi = {10.1038/s41567-018-0315-5},
   issn = {1745-2473},
   issue = {1},
   journal = {Nature Physics},
   month = {1},
   pages = {10-16},
   publisher = {Nature Publishing Group},
   title = {High-harmonic generation from solids},
   volume = {15},
   year = {2019}
}

@article{Vampa2014,
   author = {G. Vampa and C. R. McDonald and G. Orlando and D. D. Klug and P. B. Corkum and T. Brabec},
   doi = {10.1103/PhysRevLett.113.073901},
   issn = {0031-9007},
   issue = {7},
   journal = {Physical Review Letters},
   month = {8},
   pages = {073901},
   title = {Theoretical Analysis of High-Harmonic Generation in Solids},
   volume = {113},
   year = {2014}
}

@article{Vampa2015,
   author = {G. Vampa and T. J. Hammond and N. Thiré and B. E. Schmidt and F. Légaré and C. R. McDonald and T. Brabec and D. D. Klug and P. B. Corkum},
   doi = {10.1103/PhysRevLett.115.193603},
   issn = {0031-9007},
   issue = {19},
   journal = {Physical Review Letters},
   month = {11},
   pages = {193603},
   publisher = {American Physical Society},
   title = {All-Optical Reconstruction of Crystal Band Structure},
   volume = {115},
   year = {2015}
}

@article{Tancogne-Dejean2017bs,
   author = {Nicolas Tancogne-Dejean and Oliver D. Mücke and Franz X. Kärtner and Angel Rubio},
   doi = {10.1103/PhysRevLett.118.087403},
   issn = {0031-9007},
   issue = {8},
   journal = {Physical Review Letters},
   month = {2},
   pages = {087403},
   publisher = {American Physical Society},
   title = {Impact of the Electronic Band Structure in High-Harmonic Generation Spectra of Solids},
   volume = {118},
   year = {2017}
}

@article{You2017,
   author = {Yong Sing You and David A. Reis and Shambhu Ghimire},
   doi = {10.1038/nphys3955},
   issn = {17452481},
   issue = {4},
   journal = {Nat. Phys.},
   month = {4},
   pages = {345-349},
   publisher = {Nature Publishing Group},
   title = {Anisotropic high-harmonic generation in bulk crystals},
   volume = {13},
   year = {2017}
}

@article{Garcia-Cabrera2024,
author = {Garc{\'{i}}a-Cabrera, Ana and Boyero-Garc{\'{i}}a, Roberto and Zurr{\'{o}}n-Cifuentes, {\'{O}}scar and Serrano, Javier and Rom{\'{a}}n, Julio San and Plaja, Luis and Hern{\'{a}}ndez-Garc{\'{i}}a, Carlos},
doi = {10.1038/s42005-023-01511-7},
issn = {2399-3650},
journal = {Communications Physics},
month = {jan},
number = {1},
pages = {1--10},
publisher = {Nature Publishing Group},
title = {{Topological high-harmonic spectroscopy}},
volume = {7},
year = {2024}
}

@article{Ikemachi2017,
   author = {Takuya Ikemachi and Yasushi Shinohara and Takeshi Sato and Junji Yumoto and Makoto Kuwata-Gonokami and Kenichi L. Ishikawa},
   doi = {10.1103/PhysRevA.95.043416},
   issn = {2469-9926},
   issue = {4},
   journal = {Physical Review A},
   month = {4},
   pages = {043416},
   title = {Trajectory analysis of high-order-harmonic generation from periodic crystals},
   volume = {95},
   year = {2017}
}

@article{Kovalev2020,
   author = {Sergey Kovalev and Renato M.A. Dantas and Semyon Germanskiy and Jan Christoph Deinert and Bertram Green and Igor Ilyakov and Nilesh Awari and Min Chen and Mohammed Bawatna and Jiwei Ling and Faxian Xiu and Paul H.M. van Loosdrecht and Piotr Surówka and Takashi Oka and Zhe Wang},
   doi = {10.1038/s41467-020-16133-8},
   issn = {2041-1723},
   issue = {1},
   journal = {Nature Communications},
   month = {5},
   pages = {1-6},
   pmid = {32415119},
   publisher = {Nature Publishing Group},
   title = {Non-perturbative terahertz high-harmonic generation in the three-dimensional Dirac semimetal Cd3As2},
   volume = {11},
   year = {2020}
}

@article{Uzan-Narovlansky2022,
   author = {Ayelet J. Uzan-Narovlansky and Álvaro Jiménez-Galán and Gal Orenstein and Rui E.F. Silva and Talya Arusi-Parpar and Sergei Shames and Barry D. Bruner and Binghai Yan and Olga Smirnova and Misha Ivanov and Nirit Dudovich},
   doi = {10.1038/s41566-022-01010-1},
   issn = {17494893},
   issue = {6},
   journal = {Nat. Photon.},
   month = {6},
   pages = {428-432},
   publisher = {Nature Research},
   title = {Observation of light-driven band structure via multiband high-harmonic spectroscopy},
   volume = {16},
   year = {2022}
}

@article{Uzan-Narovlansky2023,
   author = {Ayelet J. Uzan-Narovlansky and Gal Orenstein and Sergei Shames and Matan Even Tzur and Omer Kneller and Barry D. Bruner and Talya Arusi-Parpar and Oren Cohen and Nirit Dudovich},
   doi = {10.1103/PhysRevLett.131.223802},
   issn = {0031-9007},
   issue = {22},
   journal = {Physical Review Letters},
   month = {11},
   pages = {223802},
   title = {Revealing the Interplay between Strong Field Selection Rules and Crystal Symmetries},
   volume = {131},
   year = {2023}
}

@article{Langer2018,
   author = {F. Langer and C. P. Schmid and S. Schlauderer and M. Gmitra and J. Fabian and P. Nagler and C. Schüller and T. Korn and P. G. Hawkins and J. T. Steiner and U. Huttner and S. W. Koch and M. Kira and R. Huber},
   doi = {10.1038/s41586-018-0013-6},
   issn = {14764687},
   issue = {7703},
   journal = {Nature},
   month = {5},
   pages = {76-80},
   pmid = {29720633},
   publisher = {Nature Publishing Group},
   title = {Lightwave valleytronics in a monolayer of tungsten diselenide},
   volume = {557},
   year = {2018}
}

@article{Tyulnev2024,
   author = {Igor Tyulnev and Álvaro Jiménez-Galán and Julita Poborska and Lenard Vamos and Philip St. J. Russell and Francesco Tani and Olga Smirnova and Misha Ivanov and Rui E. F. Silva and Jens Biegert},
   doi = {10.1038/s41586-024-07156-y},
   issn = {0028-0836},
   issue = {8009},
   journal = {Nature},
   month = {4},
   pages = {746-751},
   publisher = {Nature Publishing Group},
   title = {Valleytronics in bulk MoS2 with a topologic optical field},
   volume = {628},
   year = {2024}
}

@article{Schubert2014,
   author = {O. Schubert and M. Hohenleutner and F. Langer and B. Urbanek and C. Lange and U. Huttner and D. Golde and T. Meier and M. Kira and S. W. Koch and R. Huber},
   doi = {10.1038/nphoton.2013.349},
   issn = {1749-4885},
   issue = {2},
   journal = {Nature Photonics},
   month = {2},
   pages = {119-123},
   title = {Sub-cycle control of terahertz high-harmonic generation by dynamical Bloch oscillations},
   volume = {8},
   year = {2014}
}

@article{Uzan-Narovlansky2024,
   author = {Ayelet J. Uzan-Narovlansky and Lior Faeyrman and Graham G. Brown and Sergei Shames and Vladimir Narovlansky and Jiewen Xiao and Talya Arusi-Parpar and Omer Kneller and Barry D. Bruner and Olga Smirnova and Rui E. F. Silva and Binghai Yan and Álvaro Jiménez-Galán and Misha Ivanov and Nirit Dudovich},
   doi = {10.1038/s41586-023-06828-5},
   issn = {0028-0836},
   issue = {7997},
   journal = {Nature},
   month = {2},
   pages = {66-71},
   publisher = {Nature Publishing Group},
   title = {Observation of interband Berry phase in laser-driven crystals},
   volume = {626},
   year = {2024}
}

@article{Schmid2021,
   author = {C. P. Schmid and L. Weigl and P. Grössing and V. Junk and C. Gorini and S. Schlauderer and S. Ito and M. Meierhofer and N. Hofmann and D. Afanasiev and J. Crewse and K. A. Kokh and O. E. Tereshchenko and J. Güdde and F. Evers and J. Wilhelm and K. Richter and U. Höfer and R. Huber},
   doi = {10.1038/s41586-021-03466-7},
   issn = {0028-0836},
   issue = {7859},
   journal = {Nature},
   month = {5},
   pages = {385-390},
   publisher = {Nature Publishing Group},
   title = {Tunable non-integer high-harmonic generation in a topological insulator},
   volume = {593},
   year = {2021}
}

@article{Bai2021,
   author = {Ya Bai and Fucong Fei and Shuo Wang and Na Li and Xiaolu Li and Fengqi Song and Ruxin Li and Zhizhan Xu and Peng Liu},
   doi = {10.1038/s41567-020-01052-8},
   issn = {17452481},
   issue = {3},
   journal = {Nat. Phys.},
   month = {3},
   pages = {311-315},
   publisher = {Nature Research},
   title = {High-harmonic generation from topological surface states},
   volume = {17},
   year = {2021}
}

@article{Baykusheva2021,
    author = {Denitsa Baykusheva and Alexis Chacón and Jian Lu and Trevor P. Bailey and Jonathan A. Sobota and Hadas Soifer and Patrick S. Kirchmann and Costel Rotundu and Ctirad Uher and Tony F. Heinz and David A. Reis and Shambhu Ghimire},
   doi = {10.1021/acs.nanolett.1c02145},
   issn = {1530-6984},
   issue = {21},
   journal = {Nano Letters},
   month = {11},
   pages = {8970-8978},
   pmid = {34676752},
   publisher = {American Chemical Society},
   title = {All-Optical Probe of Three-Dimensional Topological Insulators Based on High-Harmonic Generation by Circularly Polarized Laser Fields},
   volume = {21},
   year = {2021}
}

@article{Tancogne-Dejean2017,
   author = {Nicolas Tancogne-Dejean and Oliver D. Mücke and Franz X. Kärtner and Angel Rubio},
   doi = {10.1038/s41467-017-00764-5},
   issn = {2041-1723},
   issue = {1},
   journal = {Nature Communications},
   month = {12},
   pages = {745},
   publisher = {Springer US},
   title = {Ellipticity dependence of high-harmonic generation in solids originating from coupled intraband and interband dynamics},
   volume = {8},
   year = {2017}
}

@article{Klemke2019,
   author = {N. Klemke and N. Tancogne-Dejean and G. M. Rossi and Y. Yang and F. Scheiba and R. E. Mainz and G. Di Sciacca and A. Rubio and F. X. Kärtner and O. D. Mücke},
   doi = {10.1038/s41467-019-09328-1},
   issn = {2041-1723},
   issue = {1},
   journal = {Nature Communications},
   month = {3},
   pages = {1319},
   publisher = {Nature Publishing Group},
   title = {Polarization-state-resolved high-harmonic spectroscopy of solids},
   volume = {10},
   year = {2019}
}

@article{Lv2021,
   author = {Yang-Yang Lv and Jinlong Xu and Shuang Han and Chi Zhang and Yadong Han and Jian Zhou and Shu-Hua Yao and Xiao-Ping Liu and Ming-Hui Lu and Hongming Weng and Zhenda Xie and Yan B. Chen and Jianbo Hu and Yan-Feng Chen and Shining Zhu},
   doi = {10.1038/s41467-021-26766-y},
   issn = {2041-1723},
   issue = {1},
   journal = {Nature Communications},
   month = {11},
   pages = {6437},
   pmid = {34750384},
   publisher = {Nature Research},
   title = {High-harmonic generation in Weyl semimetal $\beta$-WP2 crystals},
   volume = {12},
   year = {2021}
}

@article{Zhang2024,
   author = {Xiao Zhang and Jeroen van den Brink and Jinbin Li},
   month = {5},
   title = {Impact of crystal symmetries and Weyl nodes on high-harmonic generation in Weyl semimetal TaAs},
   doi ={10.48550/arXiv.2405.16125},
   year = {2024}
}

@article{Ominato2025,
   author = {Yuya Ominato and Masahito Mochizuki},
   doi = {10.1103/PhysRevResearch.7.023218},
   issn = {2643-1564},
   issue = {2},
   journal = {Physical Review Research},
   month = {6},
   pages = {023218},
   title = {Theory of photocurrent and high-harmonic generation with chiral fermions},
   volume = {7},
   year = {2025}
}

@article{Avetissian2022,
   author = {H K Avetissian and V N Avetisyan and B R Avchyan and G F Mkrtchian},
   doi = {10.1103/PhysRevA.106.033107},
   journal = {Physical Review A},
   pages = {33107},
   title = {High-order harmonic generation in three-dimensional Weyl semimetals with broken time-reversal symmetry},
   volume = {106},
   year = {2022}
}

@article{Liu2025,
   author = {Xiulan Liu and Lei Geng and Xiao-Shuang Kong and Jianing Zhang and Yong-Kang Fang and Liang-You Peng},
   doi = {10.1103/PhysRevB.111.184314},
   issn = {2469-9950},
   issue = {18},
   journal = {Physical Review B},
   month = {5},
   pages = {184314},
   publisher = {American Physical Society},
   title = {Theoretical investigations of high harmonic generation in the Weyl semimetal   WP 2  },
   volume = {111},
   year = {2025}
}

@article{Bharti2024,
   author = {Amar Bharti and Gopal Dixit},
   doi = {10.1063/5.0217186},
   isbn = {202408:11:34},
   journal = {Appl. Phys. Lett},
   pages = {51104},
   title = {Non-perturbative nonlinear optical responses in Weyl semimetals},
   volume = {125},
   year = {2024}
}

@article{Bharti2023,
   author = {Amar Bharti and Gopal Dixit},
   doi = {10.1103/PhysRevB.107.224308},
   issn = {2469-9950},
   issue = {22},
   journal = {Physical Review B},
   month = {6},
   pages = {224308},
   title = {Role of topological charges in the nonlinear optical response from Weyl semimetals},
   volume = {107},
   year = {2023}
}

@article{Yoshikawa2017,
   author = {Naotaka Yoshikawa and Tomohiro Tamaya and Koichiro Tanaka},
   doi = {10.1126/science.aam8861},
   issn = {10959203},
   issue = {6339},
   journal = {Science},
   month = {5},
   pages = {736-738},
   pmid = {28522530},
   publisher = {American Association for the Advancement of Science},
   title = {High-harmonic generation in graphene enhanced by elliptically polarized light excitation},
   volume = {356},
   year = {2017}
}

@article{Neufeld2019,
   author = {Ofer Neufeld and Daniel Podolsky and Oren Cohen},
   doi = {10.1038/s41467-018-07935-y},
   issn = {2041-1723},
   issue = {1},
   journal = {Nature Communications},
   month = {1},
   pages = {405},
   pmid = {30679423},
   publisher = {Nature Publishing Group},
   title = {Floquet group theory and its application to selection rules in harmonic generation},
   volume = {10},
   year = {2019}
}

@article{delasHeras2024,
   author = {Alba de las Heras and David Schmidt and Julio San Román and Javier Serrano and Jonathan Barolak and Bojana Ivanic and Cameron Clarke and Nathaniel Westlake and Daniel E. Adams and Luis Plaja and Charles G. Durfee and Carlos Hernández-García},
   doi = {10.1364/OPTICA.517702},
   issn = {2334-2536},
   issue = {8},
   journal = {Optica},
   month = {8},
   pages = {1085},
   publisher = {Optica Publishing Group},
   title = {Attosecond vortex pulse trains},
   volume = {11},
   year = {2024}
}

@article{DelasHeras2022,
   author = {Alba de las Heras and Alok Kumar Pandey and Julio San Román and Javier Serrano and Elsa Baynard and Guillaume Dovillaire and Moana Pittman and Charles G. Durfee and Luis Plaja and Sophie Kazamias and Olivier Guilbaud and Carlos Hernández-García},
   doi = {10.1364/OPTICA.442304},
   issn = {2334-2536},
   issue = {1},
   journal = {Optica},
   month = {1},
   pages = {71},
   publisher = {Optical Society of America},
   title = {Extreme-ultraviolet vector-vortex beams from high harmonic generation},
   volume = {9},
   year = {2022}
}

@article{Hickstein2015,
author = {Hickstein, Daniel D. and Dollar, Franklin J. and Grychtol, Patrik and Ellis, Jennifer L. and Knut, Ronny and Hern{\'{a}}ndez-Garc{\'{i}}a, Carlos and Zusin, Dmitriy and Gentry, Christian and Shaw, Justin M. and Fan, Tingting and Dorney, Kevin M. and Becker, Andreas and Jaro{\'{n}}-Becker, Agnieszka and Kapteyn, Henry C. and Murnane, Margaret M. and Durfee, Charles G.},
doi = {10.1038/nphoton.2015.181},
issn = {1749-4885},
journal = {Nature Photonics},
month = {nov},
number = {11},
pages = {743--750},
publisher = {Nature Publishing Group},
title = {{Non-collinear generation of angularly isolated circularly polarized high harmonics}},
volume = {9},
year = {2015}
}

@article{Huang2018,
author = {Huang, Pei-Chi and Hern{\'{a}}ndez-Garc{\'{i}}a, Carlos and Huang, Jen-Ting and Huang, Po-Yao and Lu, Chih-Hsuan and Rego, Laura and Hickstein, Daniel D. and Ellis, Jennifer L. and Jaron-Becker, Agnieszka and Becker, Andreas and Yang, Shang-Da and Durfee, Charles G. and Plaja, Luis and Kapteyn, Henry C. and Murnane, Margaret M. and Kung, A. H. and Chen, Ming-Chang},
doi = {10.1038/s41566-018-0145-0},
issn = {1749-4885},
journal = {Nature Photonics},
month = {jun},
number = {6},
pages = {349--354},
publisher = {Nature Publishing Group},
title = {{Polarization control of isolated high-harmonic pulses}},
volume = {12},
year = {2018}
}

@article{Fan2015,
author = {Fan, Tingting and Grychtol, Patrik and Knut, Ronny and Hern{\'{a}}ndez-Garc{\'{i}}a, Carlos and Hickstein, Daniel D. and Zusin, Dmitriy and Gentry, Christian and Dollar, Franklin J. and Mancuso, Christopher A. and Hogle, Craig W. and Kfir, Ofer and Legut, Dominik and Carva, Karel and Ellis, Jennifer L. and Dorney, Kevin M. and Chen, Cong and Shpyrko, Oleg G. and Fullerton, Eric E. and Cohen, Oren and Oppeneer, Peter M. and Milo{\v{s}}evi{\'{c}}, Dejan B. and Becker, Andreas and Jaro{\'{n}}-Becker, Agnieszka A. and Popmintchev, Tenio and Murnane, Margaret M. and Kapteyn, Henry C.},
doi = {10.1073/PNAS.1519666112},
issn = {0027-8424},
journal = {Proceedings of the National Academy of Sciences},
month = {nov},
number = {46},
pages = {14206--14211},
pmid = {26534992},
publisher = {National Academy of Sciences},
title = {{Bright circularly polarized soft X-ray high harmonics for X-ray magnetic circular dichroism}},
volume = {112},
year = {2015}
}

@article{Brooks2025,
  author = {Nathan J. Brooks and Alba de las Heras and Bin Wang and Iona Binnie and Javier Serrano and Julio San Román and Luis Plaja and Henry C. Kapteyn and Carlos Hernández-García and Margaret M. Murnane},
   doi = {10.1021/acsphotonics.4c01996},
   issn = {2330-4022},
   issue = {1},
   journal = {ACS Photonics},
   month = {1},
   pages = {495-504},
   publisher = {American Chemical Society},
   title = {Circularly Polarized Attosecond Pulses Enabled by an Azimuthal Phase and Polarization Grating},
   volume = {12},
   year = {2025}
}

@article{Rees2020,
   author = {Dylan Rees and Kaustuv Manna and Baozhu Lu and Takahiro Morimoto and Horst Borrmann and Claudia Felser and J. E. Moore and Darius H. Torchinsky and J. Orenstein},
   doi = {10.1126/sciadv.aba0509},
   issn = {23752548},
   issue = {29},
   journal = {Science Advances},
   month = {7},
   pmid = {32832618},
   publisher = {American Association for the Advancement of Science},
   title = {Helicity-dependent photocurrents in the chiral Weyl semimetal RhSi},
   volume = {6},
   year = {2020}
}

@article{Ni2021,
   author = {Zhuoliang Ni and K. Wang and Y. Zhang and O. Pozo and B. Xu and X. Han and K. Manna and J. Paglione and C. Felser and A. G. Grushin and F. de Juan and E. J. Mele and Liang Wu},
   doi = {10.1038/s41467-020-20408-5},
   issn = {20411723},
   issue = {1},
   journal = {Nature Communications},
   month = {12},
   pmid = {33420054},
   publisher = {Nature Research},
   title = {Giant topological longitudinal circular photo-galvanic effect in the chiral multifold semimetal CoSi},
   volume = {12},
   year = {2021}
}

@article{Neufeld2021,
   author = {Ofer Neufeld and Nicolas Tancogne-Dejean and Umberto De Giovannini and Hannes Hübener and Angel Rubio},
   doi = {10.1103/PhysRevLett.127.126601},
   issn = {0031-9007},
   issue = {12},
   journal = {Physical Review Letters},
   month = {9},
   pages = {126601},
   title = {Light-Driven Extremely Nonlinear Bulk Photogalvanic Currents},
   volume = {127},
   year = {2021}
}

@article{Ma2022,
   author = {Junchao Ma and Bin Cheng and Lin Li and Zipu Fan and Haimen Mu and Jiawei Lai and Xiaoming Song and Dehong Yang and Jinluo Cheng and Zhengfei Wang and Changgan Zeng and Dong Sun},
   doi = {10.1038/s41467-022-33190-3},
   issn = {20411723},
   issue = {1},
   journal = {Nature Communications},
   month = {12},
   pmid = {36109522},
   publisher = {Nature Research},
   title = {Unveiling Weyl-related optical responses in semiconducting tellurium by mid-infrared circular photogalvanic effect},
   volume = {13},
   year = {2022}
}

@article{Chang2017,
   author = {Guoqing Chang and Su-Yang Xu and Benjamin J. Wieder and Daniel S. Sanchez and Shin-Ming Huang and Ilya Belopolski and Tay-Rong Chang and Songtian Zhang and Arun Bansil and Hsin Lin and M. Zahid Hasan},
   doi = {10.1103/PhysRevLett.119.206401},
   issn = {0031-9007},
   issue = {20},
   journal = {Physical Review Letters},
   month = {11},
   pages = {206401},
   title = {Unconventional Chiral Fermions and Large Topological Fermi Arcs in RhSi},
   volume = {119},
   year = {2017}
}

@article{Hasan2021,
   author = {M. Zahid Hasan and Guoqing Chang and Ilya Belopolski and Guang Bian and Su-Yang Xu and Jia-Xin Yin},
   doi = {10.1038/s41578-021-00301-3},
   issn = {2058-8437},
   issue = {9},
   journal = {Nature Reviews Materials},
   month = {4},
   pages = {784-803},
   publisher = {Nature Publishing Group},
   title = {Weyl, Dirac and high-fold chiral fermions in topological quantum matter},
   volume = {6},
   year = {2021}
}

@article{Yan2024,
   author = {Binghai Yan},
   doi = {10.1146/annurev-matsci-080222-033548},
   issn = {1531-7331},
   issue = {1},
   journal = {Annual Review of Materials Research},
   month = {8},
   pages = {97-115},
   publisher = {Annual Reviews},
   title = {Structural Chirality and Electronic Chirality in Quantum Materials},
   volume = {54},
   year = {2024}
}

@article{Chen2020,
   author = {Zi Yu Chen and Rui Qin},
   doi = {10.1103/PhysRevA.101.053423},
   issn = {24699934},
   issue = {5},
   journal = {Physical Review A},
   month = {5},
   publisher = {American Physical Society},
   title = {Probing structural chirality of crystals using high-order harmonic generation in solids},
   volume = {101},
   year = {2020}
}

@article{Heinrich2021,
   author = {Tobias Heinrich and Marco Taucer and Ofer Kfir and P. B. Corkum and André Staudte and Claus Ropers and Murat Sivis},
   doi = {10.1038/s41467-021-23999-9},
   issn = {2041-1723},
   issue = {1},
   journal = {Nature Communications},
   month = {6},
   pages = {3723},
   pmid = {34140484},
   publisher = {Nature Publishing Group},
   title = {Chiral high-harmonic generation and spectroscopy on solid surfaces using polarization-tailored strong fields},
   volume = {12},
   year = {2021}
}

@article{Neufeld2019chiral,
   author = {Ofer Neufeld and David Ayuso and Piero Decleva and Misha Y. Ivanov and Olga Smirnova and Oren Cohen},
   doi = {10.1103/PHYSREVX.9.031002},
   issn = {21603308},
   issue = {3},
   journal = {Physical Review X},
   month = {7},
   pages = {031002},
   publisher = {American Physical Society},
   title = {Ultrasensitive Chiral Spectroscopy by Dynamical Symmetry Breaking in High Harmonic Generation},
   volume = {9},
   year = {2019}
}

@article{Mayer2024,
   author = {Nicola Mayer and David Ayuso and Piero Decleva and Margarita Khokhlova and Emilio Pisanty and Misha Ivanov and Olga Smirnova},
   doi = {10.1038/s41566-024-01499-8},
   issn = {17494893},
   issue = {11},
   journal = {Nature Photonics},
   month = {11},
   pages = {1155-1160},
   publisher = {Nature Research},
   title = {Chiral topological light for detection of robust enantiosensitive observables},
   volume = {18},
   year = {2024}
}

@article{Habibovi2024,
   author = {Dino Habibović and Kathryn R. Hamilton and Ofer Neufeld and Laura Rego},
   doi = {10.1038/s42254-024-00763-8},
   issn = {2522-5820},
   issue = {11},
   journal = {Nature Reviews Physics},
   month = {9},
   pages = {663-675},
   publisher = {Nature Publishing Group},
   title = {Emerging tailored light sources for studying chirality and symmetry},
   volume = {6},
   year = {2024}
}

@article{Wanie2024,
   author = {Vincent Wanie and Etienne Bloch and Erik P. Månsson and Lorenzo Colaizzi and Sergey Ryabchuk and Krishna Saraswathula and Andres F. Ordonez and David Ayuso and Olga Smirnova and Andrea Trabattoni and Valérie Blanchet and Nadia Ben Amor and Marie-Catherine Heitz and Yann Mairesse and Bernard Pons and Francesca Calegari},
   doi = {10.1038/s41586-024-07415-y},
   issn = {0028-0836},
   issue = {8015},
   journal = {Nature},
   month = {6},
   pages = {109-115},
   publisher = {Nature Publishing Group},
   title = {Capturing electron-driven chiral dynamics in UV-excited molecules},
   volume = {630},
   year = {2024}
}

@article{Li1989,
   author = {X. F. Li and A. L’Huillier and M. Ferray and L. A. Lompré and G. Mainfray},
   doi = {10.1103/PhysRevA.39.5751},
   issn = {0556-2791},
   issue = {11},
   journal = {Physical Review A},
   month = {6},
   pages = {5751-5761},
   publisher = {American Physical Society},
   title = {Multiple-harmonic generation in rare gases at high laser intensity},
   volume = {39},
   year = {1989}
}

@article{Liu2017,
   author = {Xi Liu and Xiaosong Zhu and Xiaofan Zhang and Dian Wang and Pengfei Lan and Peixiang Lu},
   doi = {10.1364/OE.25.029216},
   issn = {1094-4087},
   issue = {23},
   journal = {Optics Express},
   month = {11},
   pages = {29216},
   title = {Wavelength scaling of the cutoff energy in the solid high harmonic generation},
   volume = {25},
   year = {2017}
}

@article{Wu2015,
   author = {Mengxi Wu and Shambhu Ghimire and David A. Reis and Kenneth J. Schafer and Mette B. Gaarde},
   doi = {10.1103/PhysRevA.91.043839},
   issn = {1050-2947},
   issue = {4},
   journal = {Physical Review A},
   month = {4},
   pages = {043839},
   publisher = {American Physical Society},
   title = {High-harmonic generation from Bloch electrons in solids},
   volume = {91},
   year = {2015}
}

@article{Luu2015,
   author = {T. T. Luu and M. Garg and S. Yu. Kruchinin and A. Moulet and M. Th Hassan and E. Goulielmakis},
   doi = {10.1038/nature14456},
   issn = {14764687},
   issue = {7553},
   journal = {Nature},
   month = {5},
   pages = {498-502},
   publisher = {Nature Publishing Group},
   title = {Extreme ultraviolet high-harmonic spectroscopy of solids},
   volume = {521},
   year = {2015}
}

@article{Wu2016,
   author = {Mengxi Wu and Dana A. Browne and Kenneth J. Schafer and Mette B. Gaarde},
   doi = {10.1103/PhysRevA.94.063403},
   issn = {24699934},
   issue = {6},
   journal = {Physical Review A},
   month = {12},
   publisher = {American Physical Society},
   title = {Multilevel perspective on high-order harmonic generation in solids},
   volume = {94},
   year = {2016}
}

@article{Hawkins2015,
   author = {Peter G. Hawkins and Misha Yu. Ivanov and Vladislav S. Yakovlev},
   doi = {10.1103/PhysRevA.91.013405},
   issn = {1050-2947},
   issue = {1},
   journal = {Physical Review A},
   month = {1},
   pages = {013405},
   publisher = {American Physical Society},
   title = {Effect of multiple conduction bands on high-harmonic emission from dielectrics},
   volume = {91},
   year = {2015}
}

@article{Hentschel2001,
   author = {M. Hentschel and R. Kienberger and Ch. Spielmann and G. A. Reider and N. Milosevic and T. Brabec and P. Corkum and U. Heinzmann and M. Drescher and F. Krausz},
   doi = {10.1038/35107000},
   issn = {00280836},
   issue = {6863},
   journal = {Nature},
   month = {11},
   pages = {509-513},
   publisher = {Nature Publishing Group},
   title = {Attosecond metrology},
   volume = {414},
   year = {2001}
}

@article{Paul2001,
   author = {P. M. Paul and E. S. Toma and P. Breger and G. Mullot and F. Augé and Ph. Balcou and H. G. Muller and P. Agostini},
   doi = {10.1126/SCIENCE.1059413},
   issn = {0036-8075},
   issue = {5522},
   journal = {Science},
   month = {6},
   pages = {1689-1692},
   pmid = {11387467},
   publisher = {American Association for the Advancement of Science},
   title = {Observation of a Train of Attosecond Pulses from High Harmonic Generation},
   volume = {292},
   year = {2001}
}

@article{Allegre2025,
   author = {Hortense Allegre and Joseph J. Broughton and Tim Klee and Yan Li and Katarzyna M. Kowalczyk and Nikolas Thatte and Daniel Lim and Jon P. Marangos and Mary M. Matthews and John W. G. Tisch},
   doi = {10.1364/ol.547945},
   issn = {0146-9592},
   issue = {5},
   journal = {Optics Letters},
   month = {3},
   pages = {1492},
   pmid = {40019963},
   publisher = {Optica Publishing Group},
   title = {Extension of high-harmonic generation cutoff in solids to 50 eV using MgO},
   volume = {50},
   year = {2025}
}

@article{Ndabashimiye2016,
   author = {Georges Ndabashimiye and Shambhu Ghimire and Mengxi Wu and Dana A. Browne and Kenneth J. Schafer and Mette B. Gaarde and David A. Reis},
   doi = {10.1038/nature17660},
   issn = {0028-0836},
   issue = {7608},
   journal = {Nature},
   month = {6},
   pages = {520-523},
   publisher = {Nature Publishing Group},
   title = {Solid-state harmonics beyond the atomic limit},
   volume = {534},
   year = {2016}
}

@article{You2017MgO,
   author = {Yong Sing You and Mengxi Wu and Yanchun Yin and Andrew Chew and Xiaoming Ren and Shima Gholam-Mirzaei and Dana A. Browne and Michael Chini and Zenghu Chang and Kenneth J. Schafer and Mette B. Gaarde and Shambhu Ghimire},
   doi = {10.1364/ol.42.001816},
   issn = {0146-9592},
   issue = {9},
   journal = {Optics Letters},
   month = {5},
   pages = {1816},
   pmid = {28454168},
   publisher = {Optica Publishing Group},
   title = {Laser waveform control of extreme ultraviolet high harmonics from solids},
   volume = {42},
   year = {2017}
}

@article{Han2025,
   author = {Meng Han and Jia-Bao Ji and Alexander Blech and R. Esteban Goetz and Corbin Allison and Loren Greenman and Christiane P. Koch and Hans Jakob Wörner},
   doi = {10.1038/s41586-025-09455-4},
   issn = {0028-0836},
   issue = {8079},
   journal = {Nature},
   month = {9},
   pages = {95-100},
   publisher = {Nature Publishing Group},
   title = {Attosecond control and measurement of chiral photoionization dynamics},
   volume = {645},
   year = {2025}
}

@article{Ayuso2019,
   author = {David Ayuso and Ofer Neufeld and Andres F. Ordonez and Piero Decleva and Gavriel Lerner and Oren Cohen and Misha Ivanov and Olga Smirnova},
   doi = {10.1038/s41566-019-0531-2},
   issn = {1749-4885},
   issue = {12},
   journal = {Nature Photonics},
   month = {12},
   pages = {866-871},
   publisher = {Nature Publishing Group},
   title = {Synthetic chiral light for efficient control of chiral light–matter interaction},
   volume = {13},
   year = {2019}
}

@article{Ayuso2022,
   author = {David Ayuso and Andres F. Ordonez and Olga Smirnova},
   doi = {10.1039/D2CP01009G},
   issn = {1463-9076},
   journal = {Physical Chemistry Chemical Physics},
   title = {Ultrafast chirality: the road to efficient chiral measurements},
   year = {2022}
}

\newpage



  
  



\section*{Supplementary material}

# Supplementary Materials for Pulse-duration-sensitive high harmonics and attosecond locally-chiral light from a chiral topological Weyl semimetal

Alba de las Heras*[1], Ofer Neufeld*[2], and Angel Rubio[1,3]

[1]Max Planck Institute for the Structure and Dynamics of Matter & Center for Free-Electron Laser Science, Hamburg 22761, Germany
[2]Technion Israel Institute of Technology, Faculty of Chemistry, Haifa 3200003, Israel
[3]Center for Computational Quantum Physics, The Flatiron Institute, New York 10010, USA

*Emails: alba.de-las-heras@mpsd.mpg.de, ofern@technion.ac.il, angel.rubio@mpsd.mpg.de

**This PDF file includes:**

Methods
Complementary results about:
(i) the HHG extension with the pulse duration for a higher driving intensity
(ii) the role of spectral filtering in locally chiral HHG
(iii) locally chiral HHG driven by a left-circularly polarized driver
Figures S1 to S3

## Methods

To compute the electron dynamics, we numerically solve the time-dependent Kohn-Sham equations of time-dependent density-functional theory (TDDFT) using the Octopus open-access code[1]. We consider the adiabatic local-density approximation in the exchange-correlation functional, norm-conserving pseudopotentials, and the velocity gauge of the dipole Hamiltonian. This approach has provided accurate comparison with experimental results in solid-state HHG[2, 3]. The lattice parameters of RhSi, Si and MgO are respectively set to the experimental values 4.69, 5.44, and 4.19 Å. For the convergence of the results in RhSi and Si, we choose a $16 \times 16 \times 16$ k-point grid, spatial spacing $\Delta x = \Delta y = \Delta z = 0.40$ atomic units (a.u.), and time step $\Delta t = 0.3$ a.u. The calculations in MgO require a finer k-point sampling of the Brillouin zone 28x28x28, spacing $\Delta x = \Delta y = \Delta z = 0.30$ and time step $\Delta t = 0.1$ a.u.

The driving laser pulse is described with the vector potential

$$\mathbf{A}(t) = A_0 \sin\left(\frac{\pi t}{\tau_{\text{pulse}}}\right)^{3\left|\frac{\pi(t-t_d)}{\tau} - \frac{\pi}{2}\right|/4} \cdot \Theta\left(\frac{\tau}{2} - \left|t - \frac{\tau}{2}\right|\right) \cos(\omega_0 t)\mathbf{u_p} = A_{env}(t)\cos(\omega_0 t)\mathbf{u_p}, \tag{S1}$$

being $\mathbf{u_p}$ the unitary polarization vector, and $A_{env}(t)$ the vector potential envelope[4]. Then, the high harmonic spectrum is computed from the time-dependent current $\mathbf{j}(t) = \sum_{\nu=x,y,z} j_\nu(t)\mathbf{u}_\nu$ as

$$|\mathbf{HHG}(\omega)|^2 = \sum_{\nu=x,y,z} \left|\int \frac{dj_\nu(t)}{dt} e^{-i\omega t} A_{env}(t) dt\right|^2, \tag{S2}$$

where the envelope function is included to avoid numerical artifacts from a nonvanishing signal after the laser pulse. The time-resolved spectrograms are obtained via a Gabor analysis:

$$\mathbf{S}(\omega, t) = \int_{-\infty}^{\infty} \mathbf{a}(\Omega) \, e^{-\frac{(\Omega - \omega)^2}{2\sigma^2}} \, e^{i\Omega t} d\Omega, \tag{S3}$$

using the Fourier transform of the time derivative of the current

$$\mathbf{a}(\Omega) = \int \frac{d\mathbf{j}(t)}{dt} e^{-i\Omega t} dt, \tag{S4}$$

and setting $\sigma = 2.5\omega_0$. The synthesized electric field results from a selective filtering of specific harmonics into $\mathbf{HHG_{sel}}(\omega)$ and performing the inverse Fourier transform:

$$\mathbf{E}(t) \sim \int \mathbf{HHG_{sel}}(\omega) e^{i\omega t} d\omega. \tag{S5}$$

## HHG extension with the pulse duration for a higher driving intensity

By setting a higher driving intensity of 3 TW/cm$^2$, the HHG in RhSi and Si can be extended to higher photon energies (Fig. S1). The occupation analysis reveals a greater excitation to high conduction bands (Figs. S1E,J). The time-resolved spectrograms (Figs. S1A-C,F-H) still present a linear increase of the HHG cutoff energy with the interaction time, while showing more interference signatures due to the larger number of HHG pathways involved. High energy cutoffs of 80 and 180 eV are obtained in our calculations of RhSi and Si. However, we note that the experiments need to deal with the materials damage threshold, phase matching, absorption, and dephasing. Thus, experiments are required to check the limitations in the cutoff extensions.

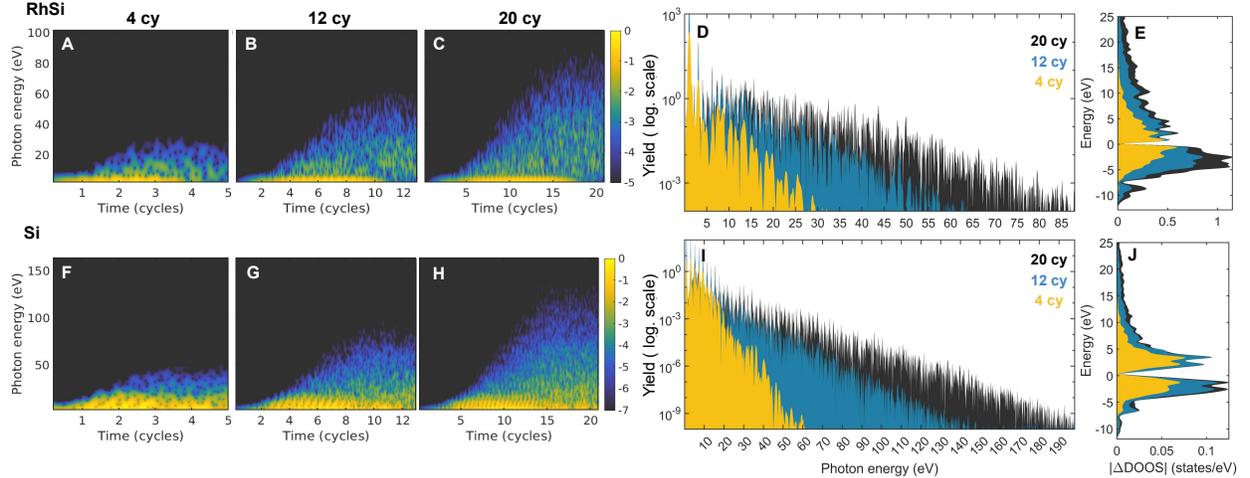

Figure S1: **Pulse duration dependence of HHG in Si and RhSi for a higher driving intensity.** Computed time-resolved spectrograms (left), high harmonic spectrum (center), and change of the density of occupied states (right).

## Role of spectral filtering in locally chiral HHG

High control in the spectral filtering can be important for the customization of locally chiral light. Fig. S2 compares the filtering of (B-D) a careful selection of multiple triplets or (E-G) beyond a threshold of $3.5\omega_0$ in the $C_3^{(+)}$ selection rule, considering a RCP driving pulse of 1 eV photon energy, 12 cycles full duration, and 1 TW/cm$^2$ peak intensity. The harmonics selected in the spectral filtering of (B-D) are indicated with

S2

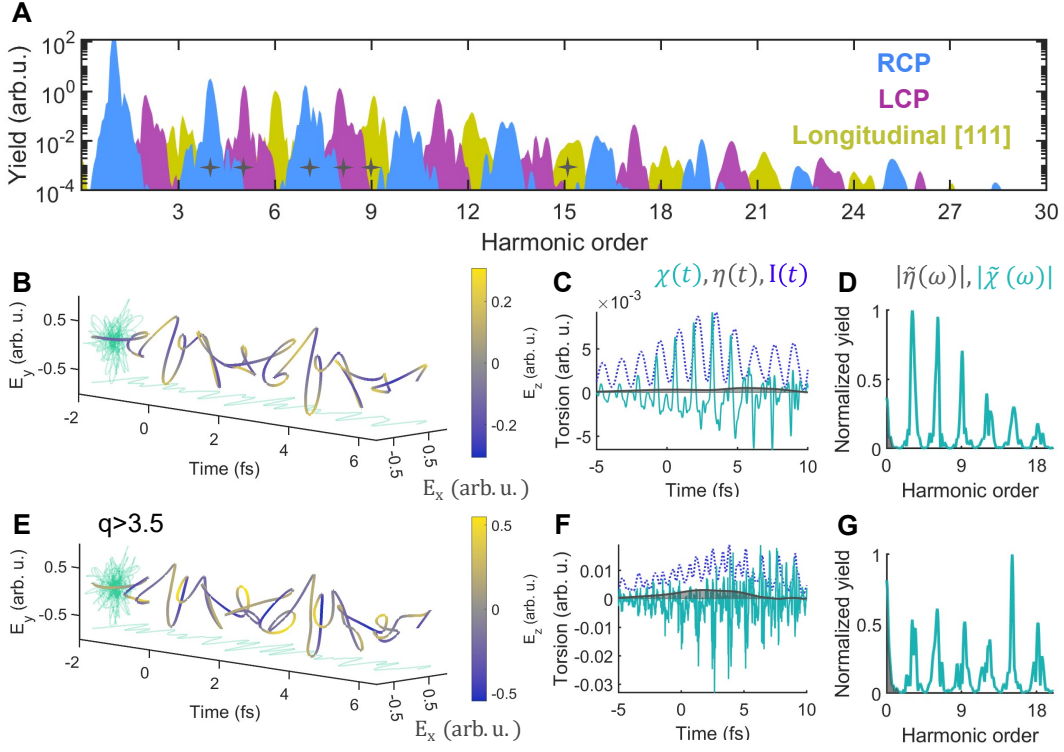

Figure S2: **Analysis of the spectral selectivity in the synthesis of attosecond locally chiral light.**
(**A**) Full high harmonic spectrum in the $C_3^{(+)}$ selection rule, considering a RCP driving pulse of 1 eV photon energy, 12 cycles full duration, and 1 TW/cm$^2$ peak intensity. (**B**) Local electric field resulting in the synthesis of starred harmonics, and its corresponding torsion in (**C**) time and (**D**) frequency domain. (**E-G**) Analogous plots are shown when filtering beyond the 3$^{\text{rd}}$ harmonic.

grey stars in (A). Whereas the 3D polarization structure of the synthesized electric field is qualitatively similar, the encoded torsion shows substantial differences. We should remark that it is possible to synthesize attosecond locally chiral light by just filtering above a harmonic threshold. However, increasing the spectral selectivity of the filter enables fine tunability of the temporal waveform of the instantaneous electric-field torsion. Figure S2(B) illustrates sharply defined torsion peaks of roughly 200 as; by contrast, filtering only above a frequency threshold leads to a significantly noisier torsion signal (Fig. S2F).

## Locally chiral HHG driven by a left-circularly polarized driver

In Fig. S3, we set the driving helicity opposite to the handedness of the chiral crystal structure. The other laser parameters are the same as Fig. 3 in the main text. The HHG presents a highly dichroic response if the driving field is contained in the (111) plane, orthogonal to the chiral crystal axis (Fig. S3A). Not only is the helicity of the harmonics reversed, but we also notice substantial differences in the harmonic spectral shape. This manifests as a strong modification in the light synthesis (Fig. S3B), where the asymmetry of the instantaneous torsion $\chi(t)$ is reduced and shows a more complex behavior (Fig. S3C). The peak of the envelope torsion in $\eta(t)$ and the net torsion $\zeta = 0.0677$ n.u. are also diminished. The differences are also manifested in the frequency domain of the torsion (Fig. S3D). On the contrary, for a driver contained in the (001) crystal plane, the high harmonic spectrum is very similar for RCP and LCP drivers, except for a flip in the helicity of the harmonics (Fig. S3E). The same torsion is obtained in Figs.3F,H, and Figs. S3F,H. For the crystal orientation (001), HHG does not sense the handedness of the crystal, but it emerges when the propagation direction is contained along the chiral crystal axis [111].

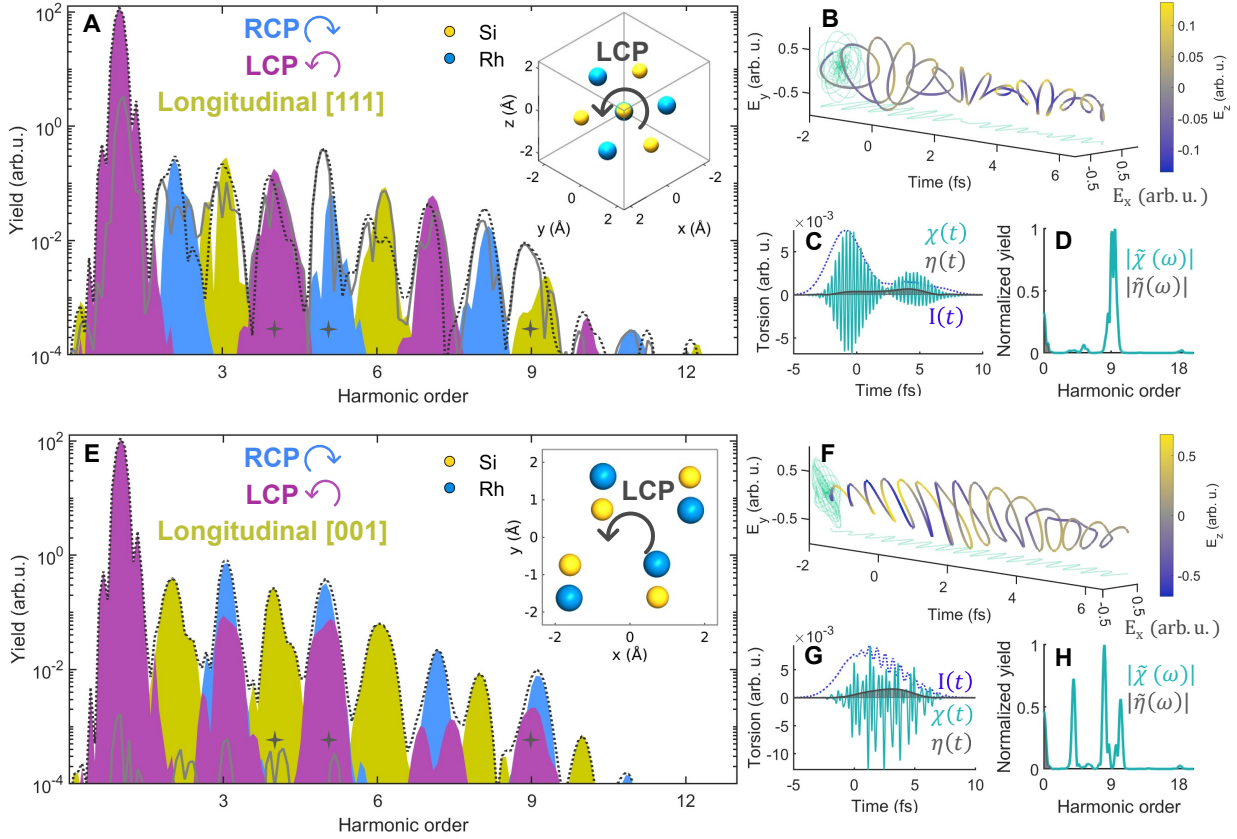

Figure S3: **Chiral high harmonic generation in RhSi driven by a LCP pulse.** Equivalent plots to Fig. 3 considering opposite driving helicity. The circular dichroism (solid line) and the case of a RCP driver (dotted line) are superimposed in (**A**) and (**E**) panels to highlight the differences in the high harmonic spectra driven from a RCP or LCP driver. The circular dichroism is maximized when the propagation direction matches the chiral axis of the crystal [111].

# References

[1] N. Tancogne-Dejean, M. J. Oliveira, X. Andrade, H. Appel, C. H. Borca, G. L. Breton, F. Buchholz, A. Castro, S. Corni, A. A. Correa, U. D. Giovannini, A. Delgado, F. G. Eich, J. Flick, G. Gil, A. Gomez, N. Helbig, H. Hübener, R. Jestädt, J. Jornet-Somoza, A. H. Larsen, I. V. Lebedeva, M. Lüders, M. A. Marques, S. T. Ohlmann, S. Pipolo, M. Rampp, C. A. Rozzi, D. A. Strubbe, S. A. Sato, C. Schäfer, I. Theophilou, A. Welden, A. Rubio, "Octopus, a computational framework for exploring light-driven phenomena and quantum dynamics in extended and finite systems", *Journal of Chemical Physics* **2020**, *152*, 124119, DOI 10.1063/1.5142502/954926.

[2] N. Tancogne-Dejean, O. D. Mücke, F. X. Kärtner, A. Rubio, "Ellipticity dependence of high-harmonic generation in solids originating from coupled intraband and interband dynamics", *Nature Communications* **2017**, *8*, 745, DOI 10.1038/s41467-017-00764-5.

[3] N. Klemke, N. Tancogne-Dejean, G. M. Rossi, Y. Yang, F. Scheiba, R. E. Mainz, G. D. Sciacca, A. Rubio, F. X. Kärtner, O. D. Mücke, "Polarization-state-resolved high-harmonic spectroscopy of solids", *Nature Communications* **2019**, *10*, 1319, DOI 10.1038/s41467-019-09328-1.

[4] O. Neufeld, O. Cohen, "Background-Free Measurement of Ring Currents by Symmetry-Breaking High-Harmonic Spectroscopy", *Physical Review Letters* **2019**, *123*, 103202, DOI 10.1103/PhysRevLett.123.103202.



\end{document}